# Teaching scenarios and their role in the interdisciplinary approach
# Case study: The Minkowskian Metric

**Ioannis Rizos**
Department of Mathematics
University of Thessaly
ioarizos@uth.gr

**Abstract**
One of the demands of modern teaching is the interdisciplinary approach of cognitive subjects and in particular that of the natural sciences, with simultaneous engagement of teachers and students. At the same time, teaching scenarios seem to be gaining ground in the methodology of teaching and learning Mathematics in school. In this paper we analyze the concepts of the teaching scenario and interdisciplinarity, and we briefly present the basic results of a teaching experiment using teaching scenario, focused on the teaching of Minkowskian Metric in two dimensions.

## 1. Introduction

It is widely accepted today that in order to understand modern mathematical concepts, we have to take seriously into account their historical evolution. Moreover, we need to consider Mathematics in interaction with other scientific fields and especially Physics. For example, one of the prominent subjects in this direction is Special Relativity Theory and its mathematical apparatus (Minkowski Geometry). This subject provides the opportunity to combine advanced concepts of Mathematics and Physics in an interdisciplinary approach.

Bearing that in mind, we will briefly describe a research of an interdisciplinary character, since it combines mathematics and science education, while at the same time borrows elements from Literature and History of Science. Parts of this research have been published in a first form in Rizos et al. (2017) and Rizos (2018).

Our main *research aim* is not to teach Special Relativity Theory, neither to introduce students into Minkowski's theory of space-time in a systematic way; our research aims at exploring high school students' conceptions about measurement of length and time, in connection with an *inquiry-based*[1] reconsideration of Geometry. More specifically, our *research question* is whether, and by what strategies, it is possible for students to use their school knowledge to work out a teaching scenario in order to formulate a mathematical model for motions and measurement in space-time.

Historical sources concerning Minkowskian Geometry and Relativity Theory are usually difficult for high school students to grasp, although interesting popularization attempts have appeared (see e.g. Eddington 1920; Gamow 1939; Taylor & Wheeler 1963; Yaglom 1979; Stannard 1989; Felsager 2004a and 2004b). Instead of using an original historical source as a text for students (see e.g. Jahnke et al. 2002) we presented to them a teaching scenario written by ourselves, which combines characters from J. Vern and H. G. Wells. This scenario was intended to lead the students to a simplified form of the Minkowskian metric, by working out some fictional numerical data. We did not impose a mathematical model to the students to deal with, but we helped them to construct it by themselves. Our didactical choice aimed to question the established epistemological conception of Geometry as an a priori system of knowledge, since a *metric* in Riemann's or Minkowski's sense could provide a "physical" meaning to mathematical concepts taught at school.

In our approach, we performed a strictly qualitative study and analyzed some limited dialogues (episodes) with a restricted number of students. They were given a teaching scenario which is presented below (see § 4). More specifically, during the school year 2013-2014 we formed a group consisting of four Greek students –two boys and two girls– who were of different educational levels (11-grade and 12-grade), with various extra-curricular activities and different family statues. In this sense, that group of students cannot be considered as "exceptional". As far as geometric axioms are concerned, the students had heard of them at school, but they were completely unaware of their role in the foundation of Geometry. In addition, all four students had meager knowledge of literature related to Science or science fiction.

---

[1] In the sense that teaching must dispute the "objectivity" and "certainty" provided by naively interpreted daily experience. More about inquiry-based teaching and learning see in Zeichner 1983; Crawford 2000 and Chapman & Heater 2010.

## 2. About teaching scenarios

In general terms, the teaching of Mathematics in school can be divided into two general categories as regards the methodology: (a) formalistic teaching using formalistic procedures, "applications", "models" and "activities" and b) teaching based on "scenarios" within a certain context. The *specifica differentia* in the methodology of teaching lies in the philosophical starting point of the two above methods. In the first case, Mathematics is self-sufficient and its teaching aims to present it to students so that they can in turn be able to *apply* it in school and in their future job. In the second case, students are invited to *interpret* a scenario within a given context or social practice and in combination with their personal experiences to "discover" the new knowledge.

Although in our days formalism –in general– is considered "undesirable" and many people give to it a negative content, nevertheless it finds the way to intrude into education (Patronis & Kapodistrias 1998). The educational context of formalism is combined with a *pragmatic view* that inspires mathematics textbooks and modern articles/ papers. According to this view, after the "theory" is taught in classroom, various practical "applications" are presented to students in order to "convince" them about the usefulness and value of Mathematics in everyday life. Thus, Mathematics is "validated" and "legalized" in students' consciousness, since its truth comes out of its own necessity.

In our research we use the teaching scenario as an *interdisciplinary context* (see Keitel 2000) and at the same time, by giving the element of dramatization, as an incentive for the students' questioning and engagement. Therefore, in Patronis' and Delikostidi's terms (2003, p. 492), we would like to define more strictly this notion: we call *teaching scenario* a finite sequence of real or imaginary actions and abstract (thought) situations, which maybe do not necessary aim to the rational achievement of a certain or obvious purpose, but they follow a pattern or they have a conceptual unit. These actions or situations are taking place in a well-known social setting/ background or in a more or less known imaginary space, that is, a familiar context for students.

The above definition differs from the current perception about scenarios in Greek secondary education. For example, in Greek School Network we can read:

«We consider as teaching scenario the description of a teaching with a focused subject(s), specific educational aims, teaching principles and practices […] In teaching scenarios elements are included such as interaction and the roles of the participants, student' perceptions and possible learning obstacles, and generally all those elements which are considered important in modern theory […] A teaching scenario is, as a rule, implemented through a series of educational activities».
"Educational material for teachers", issue 1, p. 42. Available in:
users.sch.gr/nikbalki/epim_kse/files/Parousiaseis/Edu_Scenario.pdf

Asking what are «all those elements that are considered important in modern theory» and leaving aside «educational activities», we will only comment the nebulous concept of «learning obstacles». As described in the above quotation, this concept implies a universal acceptance and objectification of the individual findings of didactical research, as if it were possible that the phenomena observed in a particular classroom were to be repeated literally in every classroom schools and at any time.

**3. Interdisciplinarity in school**
In the wide term "interdisciplinarity" we mean the joint study and exploitation of topics by different disciplines/ subjects. The concept of interdisciplinarity is intertwined with the concepts of cooperation and the spherical knowledge. In one sense, the unified teaching of Sciences is an interdisciplinary approach. For example, directly proportional quantities are today taught separately in Mathematics, Physics (rectilinear motion, uniform and uniformly accelerated), Chemistry (mass = density × volume) and elsewhere. In an interdisciplinary approach, they would teach as a single subject. According to the "Cross Curricular Thematic Framework" (C.C.T.F.) and the "Programs of Study" (Syllabi) of the former Pedagogical Institute (P.I.) and the current Institute of Educational Policy (I.E.P.) concerning Mathematics at all levels of education (Official Gazette of the Republic of Greece 303 & 304/ 13-03-03), there is an intention to enhance *cross-curricular (thematic) teaching* and *interdisciplinarity* although these terms are not clarified.[2]

However, despite the intentions of the I.E.P., what is apply in everyday school practice and which derives from the results of our research, is that very few things have finally been done in this direction. The active connection of Mathematics with the other sciences in a compatible and effective way is only left to the collaborative initiatives of some teachers, who by working silently most of the time, try to implement educational actions by applying new methodological approaches focusing on their students. On the other hand, *projects* which can theoretically provide opportunities to enhance interdisciplinarity and teamwork in teaching, are not implemented in schools by mathematicians or physicists due to… «lack of staff». But also when they are implemented, they generally end up stapling "downloaded" texts from Internet, a practice which cannot be regarded, in any sense, as a "project".

---

[2] To separate the terms *cross-curricular teaching* and *interdisciplinarity* see Beane (1997).

Furthermore, while in C.C.T.F. and "Programs of Study" (see above) references are made for «cross-curricular approaches» and «horizontal links» to each lesson separately, there is no general plan for this "interconnection". At the same time, today's secondary education lacks those courses that could combine Mathematics with Sciences and their history. For example, *Astronomy* and *History of Science and Technology* are no longer taught in secondary education. These courses could outline the evolution of the Sciences, the work and life of the people whose contribution in this evolution was remarkable. On the other hand, these subjects play an important role in preparing the ground for the students of higher secondary school to acquire fundamental concepts of Mathematics and Physics.

If the above remarks are taken into account, then it is perceived how essential it is to connect Mathematics with the other subjects (Sciences, Philosophy, History of Science, Geography, Literature, etc.) in a way that will ensure to allow students a multifaceted approach to the concepts taught. At the same time, however, each lesson should retain its autonomy otherwise it will result in a "mixture of information", inappropriate to provide solid theoretical bases to students. We consider that the courses of "interdisciplinary interest" mentioned above and the research projects can play a decisive role in this case, given the collaboration of the teachers of each field, in order to avoid the occurrence of a teacher accepting a view as true which he/ she cannot judge because of his/ her specialization.

**4. Case study: Our teaching scenario in Minkowskian metric**

In our research, we were based on a fictional scenario written by ourselves, which combines characters from J. Verne's *De la Terre à la Lune* (1865) together with the character of "Time Traveller" in H. G. Wells' *Time Machine* (1895), who appears as "*Space Traveller*" in our text, adjusted to a new situation. Our choice takes into account the fact that, as J. Verne himself used to say, his own stories were based on the actual scientific progress of his time, while the younger and talented H. G. Wells was more concerned about society in the distant future. Also, all characters of our scenario pose reasonable questions on mathematics and natural sciences and arguments to support their views.

Our text has been created with an effort to produce lively dialogues on scientific questions and an atmosphere of inquiry, contemplation and dispute. In this way, we avoided presenting scientific ideas to the readers in a "magical" way, as e.g. in the context of *Alice's adventures in Wonderland*.

Our scenario consisted of two parts: In Part I, the *Space Traveller* appeared as a young scientist who tried to verify his suspicion that Euclidean geometry cannot be the only possible geometry in physical space. The Space Traveller has access to data gathered during the travel of a spaceship, which concerns:
- the measurement of the *length* (*l*) of the spaceship and,
- the *time* (*t*) between two instantaneous gleams of a light source placed on the spaceship.

These data (see Table 1 below) have been collected separately by two groups of observers: those being in the spaceship and travelling at a speed comparable to the speed of light ("moving" observers), and those staying on the Earth ("fixed" observers). What takes place in Part II, is the mathematical treatment of the facts in Part I. Furthermore, explanations about the underlying Mathematics are given to the students.

Part II (sequel of Part I) was written by taking into account the students' responses. This part leads a mathematical elaboration and in particular the transformation of the Minkowskian metric to its quadratic form.

We must clarify here, that two dimensional Minkowski geometry is based on the (pseudo-Euclidean) quadratic form $H(x, y) = x^2 - y^2$ and its underlying space is the plane of perception, provided with the Cartesian coordinates $(x, y)$. An important example of isometric transformations in 2D Minkowski geometry is the linear transformations $x' = \alpha x + \beta y$ and $y' = \beta x + \alpha y$, where $\alpha, \beta \in \mathbb{R}$ and $\alpha^2 - \beta^2 = 1$, which are called *hyperbolic rotations*. It is notable that the area of any rectangle remains invariant under these transformations. Taking into account these facts, it can be easily obtained (Yaglom 1979, pp. 179-180) that the formula of "distance" between two vectors $\boldsymbol{x} = (x_1, y_1)$ and $\boldsymbol{y} = (x_2, y_2)$ in the Minkowskian plane is

$$d(\boldsymbol{x}, \boldsymbol{y}) = \sqrt{|(x_2 - x_1)^2 - (y_2 - y_1)^2|}$$

This metric is equivalent also to

$$d(\boldsymbol{X}, \boldsymbol{Y}) = \sqrt{2|(X_2 - X_1) \cdot (Y_2 - Y_1)|}$$

where $\boldsymbol{X} = (X_1, Y_1)$ and $\boldsymbol{Y} = (X_2, Y_2)$ are the 45° rotation of $\boldsymbol{x}$ and $\boldsymbol{y}$.

Now we can be led to the concept of (2D) space-time by making use of an informal presentation of Special Relativity by Einstein (1962). Indeed, this interpretation is achieved if we set $y = ct$, where $t$ is the time and $c$ is the speed of light, which we demand to be constant. With an appropriate choice of the above constants α, β imposed by the physical constraints, and by setting $|u| < c = 1$, we get the following well-known form of Lorentz transformations

$$x' = \frac{1}{\sqrt{1-u^2}}x - \frac{u}{\sqrt{1-u^2}}y, \quad y' = -\frac{u}{\sqrt{1-u^2}}x + \frac{1}{\sqrt{1-u^2}}y$$

The physical consequence of these fundamental transformations is *length contraction* and *time dilation* of a body moving at high speed (near the speed of light).

Below we refer only to Part I of our scenario. The text of this part is followed by a number of questions to be answered by students. Some of these questions are briefly presented below.

Question 1. *Notice the data in Table 1 about the "fixed" observers and try to think of a relation connecting the spaceship length l and the time t between two gleams of light, as the velocity v of the spaceship changes*:

*Table 1*

| v | l | t |
|---|---|---|
| 0.2 | 35.26 | 1,021 |
| 0.33 | 33,994 | 1,059 |
| 0.4 | 33 | 1,091 |
| 0.553 | 30 | 1.2 |
| 0.6 | 28.8 | 1.25 |
| 0.745 | 24 | 1.5 |
| 0.832 | 20 | 1.8 |
| 0.866 | 18 | 2 |
| 0.89 | 16,415 | 2,193 |
| 0.943 | 12 | 3 |
| 0.96 | 10 | 3.6 |
| 0.968 | 9 | 4 |

Question 2. *In an orthogonal system of coordinates place the points with coordinates (l, t) of the previous table. What can you notice? What kind of relation exists between the values of l and those of t? Is your first estimation, which you formed out of Question 1, confirmed or not?*

Question 3. *Well, the astronauts ("moving" observers) find out that the length as well as the time span remain unchangeable, but in contrast for the "fixed" observers these magnitudes bear a change. Finally, who's right? What is valid in reality? In your opinion, where is this differentiation attributed to?*

Following the length *l* of the spaceship was symbolized with *x* and the length of the time span *t* which takes place between the two gleams, with *y*. So, the use of kinematics' terms is avoided, while at the same time the same unit of measurement is attributed to these magnitudes (*x* and *y*). This unit becomes a unit of length if were place *t* with $y = c \cdot t$, so *y* will be given in $\frac{m}{sec} \cdot sec = m$. As a result, the graph of Question 2 remains the same with the only difference the renaming of the axes from (*l*, *t*) to (*x*, *y*), as follows:

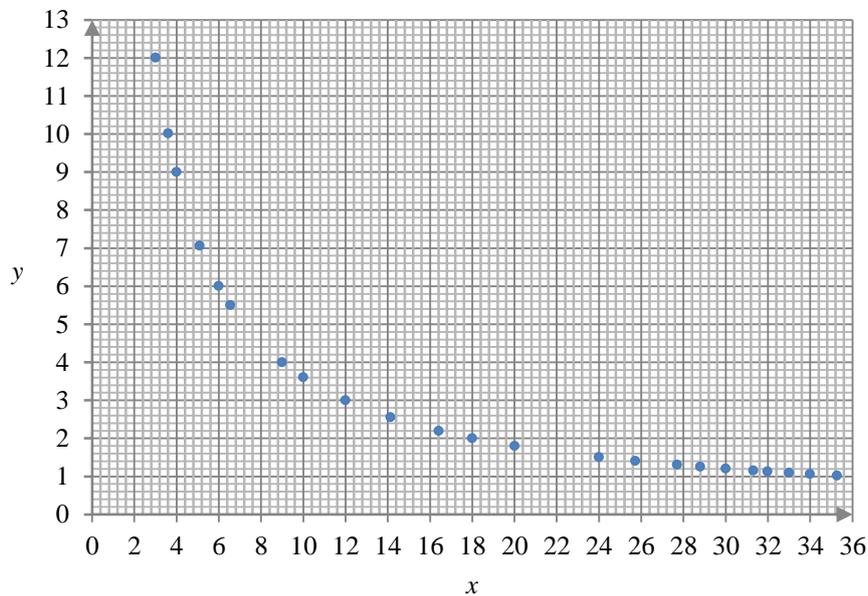

Parallel line segments were drawn from each point of the graph towards the semi-axes O*x* and O*y* which form along with the semi-axes rectangular parallelograms.

Question 4. *What do all these rectangles have in common?*

A final question was added about a controversy which was supposed to occur involving the characters of the scenario, on the rectangles formed in the fourth question: *Can these rectangles be considered as actually one and the same rectangle, in the sense that the various shapes appearing may be due to a kind of "motion" of the rectangle in the given coordinate system*? The Space Traveller made here a conjecture, advocating a view of geometry as the study of properties of figures remaining invariant under certain transformations. Here, an important such property is the *area* of rectangles. Thus, the characters of the scenario are on the threshold of a major breakthrough of a new geometry on the plane.

Question 5. *Can you find a "tape measure" according to which the length of the diagonals of the rectangles can always remain the same?*

We need here to justify our choice of providing the students with a *fictional* table of data, instead of real ones. Because of their large-scale nature, such data are not possible to be achieved in a laboratory. Moreover, with our fictional data we did not attempt to lead the students to discover a *physical law* (such as length contraction and time dilation). Our point of view is that of a *geometrical* rather than *physical* inquiry; we do not claim that our scenario, as well as our fictional data, can serve as an introduction of young students to Special Relativity Theory. Nevertheless, our fictional data are not completely arbitrary, but they are in accordance with certain predictions and thought experiments in the early development of Special Relativity (see e.g. Einstein 1962).

**5. Teaching scenarios can work – Some concluding remarks**

The difficulties encountered by the two girls who participated in the experiment can be divided into two categories: those related to the persistent conceptions of Euclidean Geometry and those due to their difficulties in the use of algebraic tools. A crucial moment for girls was when they thought the spaceship was likely to be "compressed" because of "high speed": «*is it possible that the spaceship is compressed by the high speed? But how is it* *possible that some people measure the length 36m and others as equal to a smaller one?».* The girls came to this explanation without any help except their *intuition* and *imagination*.

The two boys tended to use the Mathematics taught at school in a more "systematic" way, at least in comparison with the girls. The most important development in our research was perhaps the moment when the two boys seemed to have perceived the concept of *metric* as an additional structure in the underlying two-dimensional space. The underlying space is unique in its incidence relationships, but it can be equipped with different metrics corresponding to different geometries. In one of them terms: «*there is one geometry and in each case there is a different formula*».

In conclusion, we would say that in our research two main students' strategies were identified. One of them (belonging to the two boys) makes a creative use of preexisting school knowledge in Algebra and Geometry in order to fit the numerical data and provide the required formula. This strategy may derive from a rather formalistic school practice, but without completely missing the physical meaning. The other strategy (belonging to the two girls) initially resorts to *popular myths* (cf. Numbers & Kampourakis 2015) and then passes to an explanation relying on imagination and intuition.

If we consider the above strategies combined (if the two girls' intuitive approach harmonizes with the appropriate mathematical background with which the two boys were more familiar with), then one could say that they approached the core of the subject.

Out of the difficulties of the students we mentioned and the analysis of the dialogues of the teaching experiment, arises a need for further educational action in Mathematics in secondary education. Some possible didactic interventions that could both help overcome the above difficulties and show some prospects for Mathematical Education in Greece, are the correlation of Mathematics and its History with Literature and in general the subject of interdisciplinarity through the synthesis of teaching scenarios.

Nowadays, as it turns out in practice, two philosophical considerations for Mathematics and Mathematics Education seem to have prevailed: According to the first ("neoplatonistic") perception, Mathematics exists independently of the human activity and it is taught as a tool in order to make "how the world works" understood. According to the second (constructivistic) view, Mathematics is a cultural construction which is built within each individual person and the society as a whole, while at the same time that construction has applications in real problems.

However, Mathematics, when taught through carefully designed teaching scenarios seemed to tend to bridge the two above mentioned considerations, while simultaneously provides the perspective of studying the axiomatization of mathematical theories and the partial autonomy of mathematical objects by their "creators". Mathematics thus becomes the necessary and sufficient conceptual framework in which theories about nature can be interpreted. Of course an interdisciplinary approach is necessary that takes into account the historical evolution of mathematical concepts and combines the use of mathematical structures and models (sciences *de facto* cannot be dematheticize), with the active imagination and intuition of students. Thus, our research could be considered as a means of feasible exploitation of the reform of the Mathematics curriculum.